\documentclass[prb,twocolumn,floatfix,superscriptaddress,showpacs]{revtex4}
\usepackage{graphicx,color}
\usepackage{amsmath,amssymb,bm}

\newcommand{\mc}{\mathcal}
\newcommand{\la}{\Lambda}
\newcommand{\tn}{\textnormal}
\newcommand{\cprb}[3]{Phys.~Rev.~B {\bf #1}, #2 (#3)}
\newcommand{\cprl}[3]{Phys.~Rev.~Lett.~{\bf #1}, #2 (#3)}
\newcommand{\cnjp}[3]{New J.~Phys.~{\bf #1}, #2 (#3)}
\newcommand{\cnature}[3]{Nature {\bf #1}, #2 (#3)}
\newcommand{\cbook}[2]{\textit{#1} (#2)}

\definecolor{darkred}{rgb}{0.90,0,0}
\definecolor{darkgreen}{rgb}{0,0.60,.2}
\definecolor{darkblue}{rgb}{0,0,1}
\definecolor{grey}{cmyk}{0,0,0,0.25}
\definecolor{orange}{cmyk}{0,0.6,0.8,0}

\begin{document}
\title{Josephson current through a single Anderson impurity coupled to BCS leads}
\author{C.\ Karrasch}
\affiliation{Institut f\"ur Theoretische Physik, Universit\"at G\"ottingen, D-37077 G\"ottingen, Germany}
\author{A.\ Oguri}
\affiliation{Department of Material Science, Osaka City University, Sumiyoshi-ku, Osaka 558-8585, Japan}
\author{V.\ Meden}
\affiliation{Institut f\"ur Theoretische Physik A, RWTH Aachen, D-52056 Aachen, Germany}

\begin{abstract}
We investigate the Josephson current $\langle J(\phi)\rangle$ through a quantum dot embedded between two superconductors showing a phase difference $\phi$. The system is modeled as a single Anderson impurity coupled to BCS leads, and the functional and the numerical renormalization group frameworks are employed to treat the local Coulomb interaction $U$. We reestablish the picture of a quantum phase transition occurring if the ratio between the Kondo temperature $T_K$ and the superconducting energy gap $\Delta$ or, at appropriate $T_K/\Delta$, the phase difference $\phi$ or the impurity energy is varied. We present accurate zero- as well as finite-temperature $T$ data for the current itself, thereby settling a dispute raised about its magnitude. For small to intermediate $U$ and at $T=0$ the truncated functional renormalization group is demonstrated to produce reliable results without the need to implement demanding numerics. It thus provides a tool to extract characteristics from experimental current-voltage measurements.
\end{abstract}

\maketitle

\section{Introduction}
\label{sec:intro}
The Kondo effect\cite{kondo,hewson} and superconductivity\cite{bcs} are two of the most striking manifestations of electronic correlations in low-energy condensed matter physics. The interplay of both phenomena, e.g.\ showing up for superconducting metals containing magnetic impurities, was first studied several decades ago.\cite{oldwork1,oldwork2,oldwork3,oldwork4,oldwork5,oldwork6,oldwork7,oldwork8,oldwork9,oldwork10} The Kondo temperature $T_K$ and the superconducting gap $\Delta$ are the two competing energy scales which govern the low-energy physics of such systems. If $T_K\gg\Delta$, local magnetic moments are screened by virtue of the Kondo effect. This causes Cooper pairs to break, and the ground state becomes a Kondo rather than a BCS singlet. In the opposite limit $T_K\ll\Delta$, Kondo screening is disturbed due to the superconducting gap at the Fermi energy. The ground state describes free magnetic moments. At temperature $T=0$, a quantum phase transition from a non-magnetic singlet to a degenerate (so-called magnetic) ground state is observed if $\Delta/T_K$ increases.

In recent years, a renewed theoretical interest in the interplay between Kondo and BCS physics has developed due to the rise of nanotechnology and the associated realization of quantum dot systems connected to superconducting leads.\cite{exp1,exp2,exp3,exp4,exp5,exp6,exp7,exp8,expnovotny,exprasmussen} The microscopic parameters of such nanoscale systems (e.g.~the energy $\epsilon$ of the quantum dot) can be easily tuned, thereby allowing to study the physics in a controlled way. In general, an equilibrium current $\langle J(\phi)\rangle$ flows through such a quantum dot Josephson junction, provided that there is a finite phase difference $\phi$ between both superconductors. From the theoretical point of view, the single impurity Anderson model with BCS superconducting leads can be used to describe the low-energy physics. If the local interaction $U$ between spin up and down electrons is sufficiently large so that the impurity is singly occupied, there is again a competition between Kondo screening and the formation of Cooper pairs. The Kondo singlet ground state becomes a magnetic doublet if $\Delta/T_K$ is increased at arbitrary phase difference $\phi$. However, the critical value $U_c(\Delta)$ describing the phase boundary depends on $\phi$. Hence, a phase transition can be observed if the phase difference is varied at appropriate $\Delta/T_K$. Likewise, a transition to the singlet state is observed if the system is driven away from particle-hole symmetry by a gate voltage $\epsilon$. At the critical point, the sign of $\langle J\rangle$ changes discontinuously at $T=0$.

If one takes the limit $\Delta\to0$, the single impurity Anderson model with ordinary Fermi liquid leads is recovered. The latter is well-known to describe strongly-correlated electron behavior. Hence, reliable many-particle methods are needed for an accurate treatment of the interaction $U$ between electrons occupying the impurity. In this paper, the functional (fRG) and numerical (NRG) renormalization group schemes will be employed. We mainly focus on the zero-temperature limit but towards the end of the paper also study finite temperature effects. The NRG is a very reliable method to investigate physical properties of systems with Coulomb interaction\cite{nrg1} and thus provides a powerful tool for an unbiased calculation of the Josephson current. Unfortunately, it requires large numerical effort and in practice only small systems of high symmetry can be treated. In contrast, truncated fRG is fast, flexible, easy-to-implement, and free of numerical parameters, but by construction limited to small to intermediate interaction strength. Comparison with NRG data, however, showed that fRG correctly describes zero-temperature (i.e.\ zero-frequency) transport properties of multi-level quantum dot geometries connected to Fermi liquid leads up to fairly large $U$.\cite{dotsystems,phaselapses} In this paper we establish the accuracy of fRG in treating the Josephson problem by comparing with analytical results at $\Delta=\infty$ as well as with our NRG data.

Experimentally, it is impossible to control the phase difference $\phi$ so that only the weighted one-period average of the Josephson current can be measured. In order to extract physical properties, the experimental data needs to be fitted by a procedure which requires current-phase characteristics as an input parameter.\cite{expnovotny} This paper shows that the fRG is an accurate and fast tool to provide $\langle J(\phi)\rangle$.

As mentioned above, the fundamental physics of magnetic impurities inside a superconductor was explained decades ago.\cite{oldwork1,oldwork2,oldwork3,oldwork4,oldwork5,oldwork6,oldwork7,oldwork8,oldwork9,oldwork10} Due to the revived interest motivated by recent experiments,\cite{exp3,exp4,expnovotny,exprasmussen} the quantum dot Josephson junction has been studied using various theoretical approaches. In particular, the NRG was applied to accurately calculate the phase boundary between the singlet and doublet phase\cite{oguri1,oguri2} as well as the single-particle spectral function.\cite{oguri3} Even though the phase transition is already captured on the Hartree-Fock level\cite{ra,yo} and by perturbative approaches,\cite{vecino,novotny} there are few quantitatively reliable results for the Josephson current at $T=0$. The atomic limit $\Delta=\infty$ can be treated analytically (see Sec.~\ref{sec:comp.atom}), and Glazman and Matveev derived an expression for $\langle J(\phi)\rangle$ in the limits of $\Gamma\to 0$ and $\Delta\to 0$, respectively.\cite{gm} Quantum Monte Carlo (QMC) calculations were carried out by Siano and Egger (SE, Ref.~\onlinecite{se}), but they show significant finite-temperature effects (QMC being an inherently finite-$T$ - method). Sellier et al.~published finite-temperature data for the infinite-$U$ Anderson model obtained from the noncrossing approximation.\cite{sellier} NRG results for $\langle J(\phi)\rangle$ at arbitrary parameters were published by Choi et al.~(CLKB, Ref.~\onlinecite{choi}), but they have been criticized by SE to be inaccurate.\cite{disputechoise1,disputechoise2} The present paper settles this dispute.

We organize this paper as follows. In Sec.~\ref{sec:model} we introduce our model Hamiltonian. In Sec.~\ref{sec:methods} we briefly comment on the general ideas of the fRG and NRG. In particular, we derive the fRG flow equations for the present context. We present our results for the Josephson current in Sec.~\ref{sec:results}. Sec.~\ref{sec:comp} is devoted to an extended discussion of different theoretical approaches to the Josephson problem. In particular, we check our NRG numerics against the exact solution at $U=0$. Analytic results for $\Delta=\infty$ provide a benchmark for the fRG approximation. We comment on the dispute between CLKB and SE. In the appendices we present details of the NRG calculation, derive an exact formula for the Josephson current in terms of the self-energy, and comment on the issue of current conservation.

\section{The model}
\label{sec:model}
\begin{figure}[t]
\includegraphics[width=0.7\linewidth,clip]{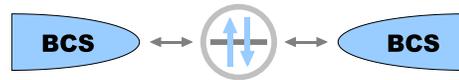}
\caption{(Color online) The quantum dot Josephson junction considered in this paper.}
\label{fig:geometry}
\end{figure}

As indicated in Fig.~\ref{fig:geometry}, our model Hamiltonian consists of three parts:
\begin{equation}\label{eq:model.h}
H = H^\tn{dot} + \sum_{s=L,R} H^\tn{lead}_s + \sum_{s=L,R}H^\tn{coup}_s.
\end{equation}
The first term describes a single Anderson impurity (the quantum dot) with onsite energy $\epsilon$ and Coulomb repulsion $U$ between spin up and spin down electrons:
\begin{equation}\begin{split}
H^\tn{dot} & = H^\tn{dot}_0+H^\tn{dot}_U \\
& = \epsilon \sum_\sigma d^\dagger_\sigma d_\sigma + U \left(d^\dagger_\uparrow d_\uparrow-\frac{1}{2}\right)\left(d^\dagger_\downarrow d_\downarrow-\frac{1}{2}\right).
\end{split}\end{equation}
The single-particle energy was shifted such that $\epsilon=0$ corresponds to the point of particle-hole symmetry. The left ($s=L$) and right ($s=R$) superconducting leads are modeled by a standard BCS Hamiltonian,
\begin{equation}
H^\tn{lead}_s = \sum_{k\sigma}\epsilon_{sk}c^\dagger_{sk\sigma}c_{sk\sigma} 
- \Delta_s\sum_k\left(e^{i\phi_s}c^\dagger_{sk\uparrow}c^\dagger_{s-k\downarrow} + \tn{H.c.}\right),
\end{equation}
where $\Delta_s$ and $\phi_s$ denote the magnitude and phase of the superconducting order parameter. $d_\sigma$ and $c_{sk\sigma}$ are the usual annihilation operators of the dot and lead electrons, respectively. The quantum dot is coupled to the leads by
\begin{equation}
H^\tn{coup}_s = -t_s\sum_\sigma\left(c^\dagger_{s\sigma}d_\sigma+\tn{H.c.}\right).
\end{equation}
The hopping matrix element $t_s$ is assumed to be real. We have introduced the local electron operator at the end of the leads, $c_{s\sigma}=\sum_k c_{sk\sigma}/\sqrt{N}$.

In order to derive the fRG flow equations, introduction of the Nambu formalism will prove to be helpful. To this end, we recast the Hamiltonian in terms of anti-commuting spinors,
\begin{equation}
\Psi_{sk} = \begin{pmatrix} c_{sk\uparrow} \\[1ex] c^\dagger_{s-k\downarrow}\end{pmatrix}~, \hspace{0.5cm}
\varphi = \begin{pmatrix} d_\uparrow \\[1ex] d^\dagger_\downarrow\end{pmatrix}.
\end{equation}
The dot part can then be written as
\begin{equation}
H^\tn{dot}  =  \epsilon\varphi^\dagger\tau_3\varphi - U \left(\varphi^\dagger_1\varphi_1-\frac{1}{2}\right)\left(\varphi^\dagger_2\varphi_2-\frac{1}{2}\right),
\end{equation}
with $\varphi_{1,2}$ denoting the components of the Nambu operator $\varphi$. We have introduced the Pauli matrices $\tau_i$. For the BCS leads we obtain the usual expression
\begin{equation}
H^\tn{lead}_s = \sum_k\left(\epsilon_{sk}\Psi_{sk}^\dagger\tau_3\Psi_{sk} - \Psi_{sk}^\dagger\bar{\Delta}_s\Psi_{sk}\right),
\end{equation}
with
\begin{equation}
\bar\Delta_{s}=\Delta_s\begin{pmatrix} 0 & e^{i\phi_s} \\ e^{-i\phi_s} & 0 \end{pmatrix}.
\end{equation}
Finally, the coupling Hamiltonian becomes
\begin{equation}
H^\tn{coup}_s = -t_s \Psi_s^\dagger\tau_3\varphi + \tn{H.c.}~,
\end{equation}
where $\Psi_s=\sum_k\Psi_{sk}/\sqrt{N}$. One should note that by introduction of the Nambu formalism, we get rid of all anomalous expectation values but have to deal with an additional single-particle quantum number (the Nambu index).

\section{The methods}
\label{sec:methods}

\subsection{Functional RG}
\label{sec:methods.frg}
The fRG is one implementation of the general renormalization group idea for interacting quantum many-particle systems.\cite{salmhofer} It starts with introducing an energy cutoff $\la$ into the noninteracting Green function $\mc G^0$. Here, we choose an infrared cutoff in Matsubara frequency space,
\begin{equation}
\mc G^0(i\omega) \longrightarrow G^{0,\la}(i\omega) = \Theta (|\omega|-\la)\mc G^0(i\omega).
\end{equation}
Using this propagator, the $m$ - particle vertex functions $\gamma_m^\la$ acquire a $\la$ - dependence. Differentiating each $\gamma_m^\la$ with respect to $\la$ yields an infinite hierarchy of flow equations
\begin{equation}\label{eq:methods.frg.floweq}
\partial_\la\gamma_m^\la=f_m\left(\left\{\gamma_n^\la\right\}\right).
\end{equation}
The functions $f_m$ can be computed straight-forwardly by introducing a generating functional for $\gamma_m$. At $\la=0$, the original cutoff-free problem is recovered. Hence, one can obtain an exact expression for the self-energy $\Sigma=-\gamma_1^{\la=0}$ by solving the set of coupled differential equations (\ref{eq:methods.frg.floweq}). In practice, this infinite hierarchy has to be truncated, thereby rendering the fRG an approximate method. In this paper, we will employ a truncation scheme that keeps the flow equations for the self-energy and the two-particle vertex $\gamma_2^\la$ evaluated at zero external frequency. The resulting approximation to the self-energy is frequency-independent and can be computed with minor numerical effort. By construction, this truncated fRG becomes exact in the noninteracting limit. One can show that it keeps track of all terms of first order in $U$ but the RG procedure leads to an efficient resummation of higher order terms. The truncated fRG was demonstrated to successfully describe strong correlation effects (such as important aspects of Kondo physics).\cite{dotsystems} 

At zero temperature, the approximate flow equations for $\Sigma^\la:=-\gamma_1^\la$ and $\Gamma^\la:=\gamma_2^\la$ explicitly read (a detailed derivation can be found in Ref.~\onlinecite{notesvolker})
\begin{equation}\label{eq:methods.frg.flowse}
\partial_\la\Sigma_{1',1}^\la = 
-\frac{1}{2\pi}\sum_{\omega=\pm\la}\sum_{2,2'}e^{i\omega\eta}
\tilde{\mc G}^\la_{2,2'}(i\omega)\Gamma_{1',2';1,2}^\la
\end{equation}
for the self-energy, and
\begin{equation}\label{eq:methods.frg.flowga}\begin{split}
\partial_\la\Gamma_{1',2';1,2}^\la = & \frac{1}{4\pi}\sum_{\omega=\pm\la}\sum_{3,3'}\sum_{4,4'}\Bigg\{
\tilde{\mc G}_{3,3'}^\la(i\omega)\tilde{\mc G}_{4,4'}^\la(-i\omega) \\[1ex]
& \times \Gamma_{1',2';3,4}^\la\Gamma_{3',4';1,2}^\la
+ 2\tilde{\mc G}_{3,3'}^\la(i\omega)\tilde{\mc G}_{4,4'}^\la(i\omega)\\[1ex]
& \times \Big[\Gamma_{2',4';1,3}^\la\Gamma_{3',1';4,2}^\la-\Gamma_{1',4';1,3}^\la\Gamma_{3',2';4,2}^\la\Big]\Bigg\}
\end{split}\end{equation}
for the effective interaction. The lower indices denote arbitrary single-particle quantum numbers. We have defined
\begin{equation}
\big[\tilde{\mc G}^\la(i\omega)\big]^{-1} = \left[\mc G^0(i\omega)\right]^{-1}-\Sigma^\la.
\end{equation}
One should note that the sharp cutoff has completely disappeared from the zero-temperature flow equations (\ref{eq:methods.frg.flowse}) and (\ref{eq:methods.frg.flowga}), rendering them easy to integrate numerically. In practice, it is convenient to exploit symmetries of the two-particle vertex (such as anti-symmetry and spin conservation) in order to reduce the number of independent equations.

Due to the static (frequency-independent) approximation involved in setting up our fRG scheme one cannot expect to obtain reliable results at finite temperatures for the problem at hand. Thus, we will focus exclusively on performing fRG calculations at $T=0$, even though the flow equations (\ref{eq:methods.frg.flowse}) and (\ref{eq:methods.frg.flowga}) can be generalized straight-forwardly to the $T>0$ - case. Studying finite-temperature effects with this method requires a more sophisticated truncation scheme and is subject of ongoing research.

A numerical solution of the flow equations starts at some large but finite $\la_0$. Due the slow decay of the r.h.s. of Eq.~(\ref{eq:methods.frg.flowse}), the integration from $\la=\infty$ down to $\la=\la_0$ always yields a finite contribution. This contribution tends to a constant if $\la_0\to\infty$ and can be computed analytically, leading to the following initial conditions at some large but finite $\la_0$:
\begin{equation}
\Sigma_{1,1'}^{\la_0\to\infty} = \frac{1}{2}\sum_2\bar{v}_{1',2;1,2}~, ~~~~
\Gamma^{\la_0\to\infty}_{1',2';1,2}=\bar{v}_{1',2';1,2},
\end{equation}
where $\bar{v}$ is the bare antisymmetrized two-particle interaction. The convergence factor in Eq.~(\ref{eq:methods.frg.flowse}) can then be dropped and the flow equations integrated numerically.

Application of the fRG scheme to the quantum dot Josephson junction is achieved straight-forwardly within the Nambu formalism. The first step consists of integrating out the noninteracting leads using a standard projection technique.\cite{taylor} Thereafter, instead of dealing with an infinite system we only have to consider two interacting (Nambu) particles. For the noninteracting dot Green function we obtain
\begin{equation}
\left[\mc G^0(i\omega)\right]^{-1} = i\omega-h^\tn{dot}_0 - \sum_{s=L,R}t_s^2\tau_3g^s(i\omega)\tau_3,
\end{equation}
where $h^\tn{dot}_0$ is the single-particle version of $H^\tn{dot}_0$ and $g^s(z)$ denotes the local propagator at the last site of the isolated BCS lead:
\begin{equation}
g^s(i\omega) = -\frac{\pi\rho_s}{\sqrt{\omega^2+\Delta_s^2}}
\begin{pmatrix}
i\omega & -\Delta_s e^{i\phi_s} \\ - \Delta_se^{-i\phi_s} & i\omega
\end{pmatrix}.
\end{equation}
We assume the local density of states at the end of the superconducting leads $\rho_s$ to be energy-independent (wide-band limit). The dot propagator then explicitly reads
\begin{equation}\label{eq:methods.frg.g0}
\left[\mc G^0(i\omega)\right]^{-1} = 
\begin{pmatrix}
i\tilde\omega - \epsilon & \tilde\Delta(i\omega) \\ \tilde\Delta(i\omega)^* & i\tilde\omega + \epsilon
\end{pmatrix},
\end{equation}
where we have introduced
\begin{equation}
i\tilde\omega = i\omega\left[1+\sum_s\frac{\Gamma_s}{\sqrt{\omega^2+\Delta_s^2}}\right],
\end{equation}
with $\Gamma_s=\pi\rho_s|t_s|^2$ being the dot-lead hybridization, and
\begin{equation}
\tilde\Delta(i\omega) = \sum_s\tilde\Delta_s(i\omega) = \sum_s\frac{\Gamma_s\Delta_s}{\sqrt{\omega^2+\Delta_s^2}}e^{i\phi_s}.
\end{equation}
In this paper we will mainly focus on the case of symmetric couplings ($\Gamma_L=\Gamma_R=\Gamma/2$) and equal superconducting gaps ($\Delta_L=\Delta_R=\Delta$). If (without loss of generality) we additionally choose $\phi_L=-\phi_R=\phi/2$, the function $\tilde\Delta(i\omega)=\Gamma\Delta\cos(\phi/2)/\sqrt{\omega^2+\Delta^2}$ becomes purely real. In general, the self-energy of the quantum dot Josephson junction is characterized by a diagonal component $\Sigma\in\mathbb{R}$ and an off-diagonal (anomalous) part $\Sigma_\Delta\in\mathbb{C}$. Within our truncated fRG approximation, they are both frequency-independent. Thus, the matrix $\tilde{\mc G}^\la$ reads
\begin{equation}
\tilde{\mc G}^\la(i\omega) = -\frac{1}{D^\la(i\omega)}
\begin{pmatrix}
i\tilde\omega + \epsilon + \Sigma^\la & \Sigma_\Delta^\la -\tilde\Delta(i\omega) \\
(\Sigma_\Delta^\la)^* -\tilde\Delta(i\omega)^* & i\tilde\omega - \epsilon - \Sigma^\la
\end{pmatrix},
\end{equation}
where we have defined the determinant 
\begin{equation}
D^\la(i\omega) = \tilde\omega^2 +\left(\epsilon+\Sigma^\la\right)^2+\left|\tilde\Delta(i\omega) - \Sigma_\Delta^\la\right|^2.
\end{equation}
The flow equations (\ref{eq:methods.frg.flowse}) for the self-energy can now be written explicitly as
\begin{equation}\label{eq:methods.frg.flowS}
\partial_\la\Sigma^\la = \frac{U^\la}{\pi}\frac{\Sigma^\la+\epsilon}{D^\la(i\la)}
\end{equation}
for the diagonal component, and
\begin{equation}\label{eq:methods.frg.flowSD}
\partial_\la\Sigma_\Delta^\la = \frac{U^\la}{\pi}\frac{\Sigma_\Delta^\la-\tilde\Delta(i\la)}{D^\la(i\la)}
\end{equation}
for the anomalous part. In the symmetric case, $\Sigma_\Delta^\la$ is real (since $\tilde\Delta$ is real). $U^\la$ denotes the only independent component of the two-particle vertex. Its flow equation (\ref{eq:methods.frg.flowga}) is given by
\begin{equation}\label{eq:methods.frg.flowU}
\partial_\la U^\la = \frac{2}{\pi}\left[\frac{U^\la}{D^\la(i\la)}\right]^2
\left[\left(\epsilon+\Sigma^\la\right)^2+\left|\Sigma_\Delta^\la-\tilde\Delta(i\la)\right|^2\right].
\end{equation}
An even simpler approximation can be obtained by neglecting the flow of the two-particle vertex altogether, using a constant interaction $U^\la=U$. While this does not affect our results qualitatively, the accuracy of fRG significantly improves if the flow of $U^\la$ is accounted for (see Sec.~\ref{sec:results.phasediag}).
The initial conditions read
\begin{equation}
\Sigma^{\la_0\to\infty}=0~,~~~ \Sigma_\Delta^{\la_0\to\infty}=0~,~~~ U^{\la_0\to\infty}=U~.
\end{equation}
In order to obtain the fRG approximation to the self-energy $\Sigma=\Sigma^{\la=0}$ and $\Sigma_\Delta=\Sigma_\Delta^{\la=0}$, we solve these coupled differential equations by numerically integrating from $\la_0=10^8$ down to $\la=0$ using standard routines.

Having obtained the self-energy of the system, the Josephson current at $T=0$ can be computed from the following integral (here focusing on $\Delta_L=\Delta_R=\Delta$, $\Gamma_L=\Gamma_R$, and $\phi_L=-\phi_R=\phi/2$):
\begin{equation}\label{eq:methods.frg.current}
\langle J \rangle =
\frac{1}{2\pi}\int\left[\frac{\Gamma^2\Delta^2\sin(\phi)}{D(i\omega)\left(\omega^2+\Delta^2\right)} 
- \frac{2\Gamma\Delta\Sigma_\Delta\sin(\phi/2)}{D(i\omega)\sqrt{\omega^2+\Delta^2}}\right]\, d\omega,
\end{equation}
where $D(i\omega)=D^{\la=0}(i\omega)$. We take $\hbar=1$ and $e=1$ (the latter being the electron charge) in the following. We will derive Eq.~(\ref{eq:methods.frg.current}) and its generalization for $T\geq0$, non-symmetric leads and a self-energy which is explicitly frequency-dependent in Appendix \ref{sec:currentformula}. In general, current conservation $\langle J_L \rangle = - \langle J_R \rangle$ is ensured for $\Delta_L=\Delta_R$ (and otherwise arbitrary parameters) within our fRG approximation. The general issue of current conservation is commented on in Appendix \ref{sec:currentconservation}.

\subsection{Numerical RG}
\label{sec:methods.nrg}
The main idea of the NRG in application to quantum impurity systems is to discretize the flat conduction band of the leads using a set of logarithmic energies $\{\pm D\la^{-n},n\geq0\}$, with $D$ being half the bandwidth and $\la>1$ the NRG discretization parameter.\cite{hewson,nrg1,thomasnrg} The resulting discrete model can be mapped onto a semi-infinite tight-binding chain with the impurity being the first site. The Hamiltonian of the semi-infinite chain is then diagonalized iteratively by adding one site at a time, starting out with the isolated impurity. Due to the logarithmic discretization, the hopping matrix elements $u_n$ between successive sites fall off exponentially with the distance $n$ from the impurity ($u_n\sim\la^{-n/2}$), rendering it possible to resolve smaller and smaller energy scales during the iteration. For a numerical implementation, a truncation procedure has to be employed as the dimension of the Hilbert space grows exponentially with the length of the chain. A very simple truncation scheme is to retain only the $N_c$ lowest-lying many-particle states at each iterative step, which is reasonable since the states of the shorter chain affect those of the longer only in a small energy window $\sim\la^{-1/2}$.

The essential approximations in the NRG approach are the logarithmic discretization of the conduction band and the truncation of the Hilbert space during the iterative diagonalization, implying that the accuracy of this method is controlled by the parameters $\la$ and $N_c$. For models containing only a single conduction band (so-called single-channel models), NRG calculations at $\la=2$ and $N_c=500$ typically agree well with analytical (such as Bethe ansatz) results. In this paper, we are considering a two-channel model which could cause problems as the number of low-energy states increases exponentially with the number of channels. Hence, NRG calculations have to be performed with care and checks at different $\la$ and $N_c$ against analytical results (available at $U=0$) are imperative (see Sec.~\ref{sec:comp.noninteracting}).

The NRG in the present work was implemented such that two sites from different channels were simultaneously added at each iteration, and we have kept the lowest $N_c=700$ energy states after the diagonalization procedure. Then the dimension of the Hamiltonian matrix in each recursive NRG step becomes approximately $700 \times 4^2$. The effective dimension in actual calculations has been reduced efficiently by the use of symmetries, which is helpful for improving the numerical accuracy. Namely, in the case of equal gaps ($\Delta_L=\Delta_R$) and symmetric couplings ($\Gamma_L=\Gamma_R$), the Hamiltonian defined in Eq.~(\ref{eq:model.h}) can be described by a real symmetric matrix even if the phase difference $\phi$ is finite (see Appendix \ref{sec:NRG_appendix}). Furthermore, in the particle-hole symmetric case ($\epsilon=0$), there is an additional U(1) symmetry associated with the conservation of the pseudo-spin variable $Q^\tn{ps}$ [Eq.~(\ref{eq:I_x^sp})]. We have carried out our NRG calculations using the quantum numbers $Q$ and $S$. The former is related to $Q^\tn{ps}$ whereas the latter corresponds to the SU(2) symmetry of the real spin. We have checked whether $N_c=700$ is enough for obtaining accurate results by increasing the number of kept states up to $N_c=1716$. It turned out that in contrast to single-channel models in our case it is not sufficient to retain only $N_c=700$ states at $\la=2$, particularly for $\Gamma\gg\Delta$. We have also confirmed that ground-state properties can be obtained reasonably well with $N_c=700$ kept states for $\Lambda \gtrsim 4$ (for the details see Sec.~\ref{sec:comp.noninteracting}). The addition of a new site during the iteration procedure can be viewed as a perturbation of relative strength of the order of $\la^{-1/2}$ (specifically for normal leads $\Delta=0$), implying that with decreasing $\la$ one has to keep more and more states to get reliable results. Hence, the numerical accuracy can be improved by increasing $\la$ at fixed $N_c$, even though the approximation involved becomes exact in the opposite limit $\la\to1$. For $\Gamma \lesssim \Delta$, the wave function of the Andreev bound state is localized well at the impurity site and does not penetrate deep inside the BCS leads. In such cases the convergence becomes better, and $N_c$ can be smaller than the value one needs for achieving the same accuracy in the opposite case $\Gamma \gg \Delta$. We note that in our calculations the NRG Hamiltonian given in Eq.~(\ref{eq:H_NRG}) is diagonalized exactly (without introducing the truncation) up to $7$ sites, which consist of the dot at the center, and the first three orbitals ($n=0,1,2$) of the $s_n$ and $p_n$ particles, respectively (see Appendix \ref{sec:NRG_appendix}). Therefore, the truncation only starts at the next NRG step for the cluster with $9$ sites.

For the NRG calculations shown in Fig.~\ref{fig:currentcomp}(a) and Fig.~\ref{fig:nrg}, we have chosen the half bandwidth such that $\Gamma=0.03847D$. We have used a different parameter for Fig.~\ref{fig:currentcomp}(b), namely $\Gamma=0.002D$. The flow of low-energy eigenvalues converges after at most $50$ NRG steps, and thus the iterations up to $n\lesssim 50$ were enough for the parameter sets we have examined. The current was computed as the expectation value of the discretized current operator given in Eq.~(\ref{eq:J_sp}).

\begin{figure}[tb]
   \includegraphics[height=5.8cm,clip]{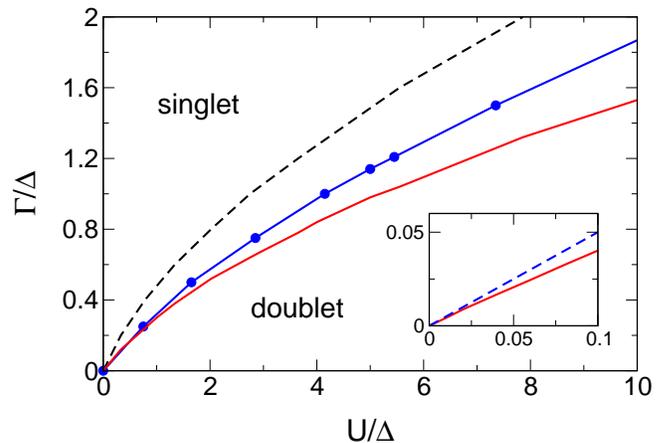}
   \caption{(Color online) The phase diagram at $\epsilon=0$, $\phi=0$, and $\Gamma_L=\Gamma_R$ characterized by the boundary line between the singlet (upper region) and doublet (lower region) phase. Solid (dashed) lines show fRG results with (without) flow of the static vertex obtained from Eqs.~(\ref{eq:methods.frg.flowS}), (\ref{eq:methods.frg.flowSD}), (\ref{eq:methods.frg.flowU}), and (\ref{eq:methods.frg.current}) and the definition of the phases via $J(\phi\to0)\gtrless0$. Symbols are NRG data (taken from Refs.~\onlinecite{oguri1} and \onlinecite{oguri2}). The inset shows that near the origin (at large $\Delta$), the slope of the phase boundary is $1/2$ (dashed line) in accordance with analytical results obtained in the atomic limit [see Eq.~(\ref{eq:comp.atom.phaseboundary})].}
   \label{fig:phasediag1}
\end{figure}

\begin{figure*}[t]
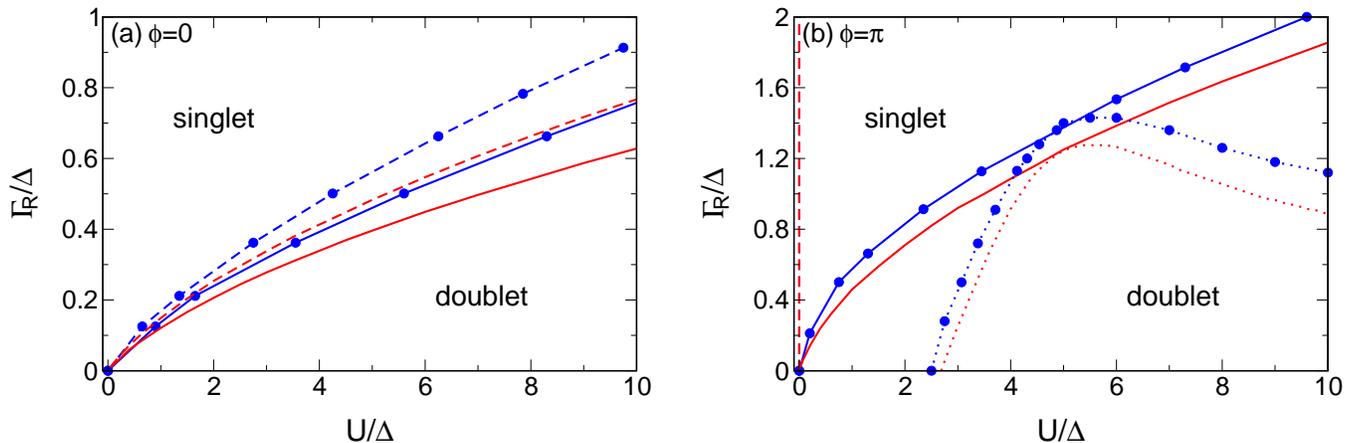

   \includegraphics[height=5.8cm,clip]{phasediag2.eps}\hspace*{0.03\linewidth}
   \includegraphics[height=5.8cm,clip]{phasediag3.eps}
   \caption{(Color online) The same as in Fig.~\ref{fig:phasediag1}, but for both symmetric $\Gamma_L=\Gamma_R$ (dashed lines) and asymmetric $\Gamma_L=1.44\Gamma_R$ (solid lines) at (a) $\phi\to0$ and (b) $\phi\to\pi$. Symbols denote NRG data. The dotted lines are the phase boundaries for $\epsilon=-2.5\Delta+U/2$, $\Gamma_L=1.44\Gamma_R$, and $\phi=\pi$.}
   \label{fig:phasediag2}
\end{figure*}

\section{Numerical Results at \boldmath$T=0$}
\label{sec:results}
In this section, we present our numerical results. Within fRG, the singlet and doublet phase is defined via $\langle J\rangle>0$ and $\langle J\rangle<0$ (focusing on $0<\phi<\pi$), respectively. Within the NRG framework, the degeneracy of the ground state is directly accessible. First, we report on what happens during the integration of the fRG flow equations. In particular, we show that the current is fairly sinusoidal in the doublet phase. Next, we show phase diagrams at ($\epsilon=0$) and away from ($\epsilon\neq0$) particle-hole symmetry which we obtain using fRG and NRG. The NRG data was previously published by one of us (A.O., Refs.~\onlinecite{oguri1} and \onlinecite{oguri2}). We stick to equal superconducting gaps ($\Delta_L=\Delta_R=\Delta$) but study both symmetric ($\Gamma_L=\Gamma_R$) and asymmetric ($\Gamma_L\neq\Gamma_R$) couplings. As all physical quantities (such as $\langle J\rangle$) depend on the phase difference only, we choose $\phi_L=-\phi_R=\phi/2$ from now on. We present new fRG and NRG results for the Josephson current $J(\phi)=\langle J(\phi)\rangle$ and demonstrate that for weak to intermediate interactions fRG performs well in comparison with NRG in describing both the phase boundary and the magnitude of $J$.

\subsection{Integrating the fRG flow equations}
\label{sec:results.frg}
First, we give a brief overview of what is happening during the integration of the fRG flow equations. For simplicity, we stick to $\Gamma_L=\Gamma_R=\Gamma/2$. In the singlet phase, nothing particular happens when (\ref{eq:methods.frg.flowS}), (\ref{eq:methods.frg.flowSD}), and (\ref{eq:methods.frg.flowU}) are solved numerically. In the doublet phase, one observes that at a certain $\la_c$, $\epsilon+\Sigma^{\la_c}$ becomes zero, implying that $\Sigma^\la=-\epsilon$ for all $\la<\la_c$ [see Eq.~(\ref{eq:methods.frg.flowS})]. The anomalous component continues to flow and reaches $\Sigma_\Delta=\Gamma\cos(\phi/2)$ at $\la=0$. Hence, the doublet phase can already be identified during the flow. Plugging $\Sigma=-\epsilon$ and $\Sigma_\Delta=\Gamma\cos(\phi/2)$ into Eq.~(\ref{eq:methods.frg.current}) yields an analytic expression for the Josephson current in the doublet phase:
\begin{equation}\label{eq:results.flow.currentpi}
J =
\int\frac{\Gamma^2\Delta\sin(\phi)}{2\pi D_d(i\omega)}\left[\frac{\Delta}{\omega^2+\Delta^2} - \frac{1}{\sqrt{\omega^2+\Delta^2}}\right] d\omega,
\end{equation}
which is a universal curve independent of both $U$ and $\epsilon$. One should note that $\sin(\phi)$ is not the only $\phi$ - dependence as
\begin{equation}
D_d(i\omega) = \tilde\omega^2+\Gamma^2\cos^2(\phi/2)\left(1-\frac{\Delta}{\sqrt{\omega^2+\Delta^2}}\right)^2.
\end{equation}
Even though the complete independence of $U$ and $\epsilon$ is an artifact of the fRG approximation, it is instructive to gain analytical insight into the current described by Eq.~(\ref{eq:results.flow.currentpi}). To this end, we scale $\Delta$ out of the integrand:
\begin{equation}
J = \int\frac{-\Gamma^2/(2\pi\Delta)\sin(\phi)\left(y-y^2\right)}{x^2(1+y(\Gamma/\Delta))^2+(\Gamma/\Delta)^2\cos^2(\phi/2)(1-y)^2}\,dx,
\end{equation}
with $y=1/\sqrt{1+x^2}$. This integral is dominated by the behavior at small $|x|$. The term proportional to $\cos(\phi/2)$ in the denominator is then of order $x^4$ and can be neglected compared to the $x^2$ term, provided that $\Gamma/\Delta$ is not too large (which is generally the case in the doublet phase). This gives
\begin{equation}\label{eq:results.flow.current}\begin{split}
\lim_{\Gamma/\Delta\to0} J & = -\frac{\Gamma^2}{2\pi\Delta}\int\frac{1}{x^2}\left(\frac{1}{\sqrt{1+x^2}}-\frac{1}{1+x^2}\right)\,dx\\
& = -\frac{\Gamma^2}{\Delta}\frac{\pi-2}{2\pi}\sin(\phi)\approx-0.18\frac{\Gamma^2}{\Delta}\sin(\phi).
\end{split}\end{equation}
Thus, the current in the doublet phase obtained from fRG is fairly sinusoidal (becoming perfectly sinusoidal with decreasing $\Gamma/\Delta$) and of order $\Gamma^2/\Delta$ (this is illustrated by Fig.~\ref{fig:atom2}; see below).

Within fRG, we compute the average occupation number $n_d$ of the quantum dot from integrating the Green function $\mc G_{1,1}=\mc G_{1,1}^{\la=0}$ over the imaginary axis:
\begin{equation}
n_d = \frac{1}{2\pi}\int e^{i\omega\eta}\mc G_{1,1}(i\omega)\,d\omega.
\end{equation}
In the doublet phase, $\mc G_{1,1}(i\omega)$ becomes an odd function (since $\Sigma+\epsilon=0$), implying that $n=1/2$ due to the contribution from large frequencies. It is again an artifact of the truncated fRG that in this phase the average occupation is completely independent of $U$ and $\epsilon$. However, NRG computations showed that the deviations from $n_d(\epsilon)=1/2$ are small.\cite{oguri2}

\begin{figure*}[t]
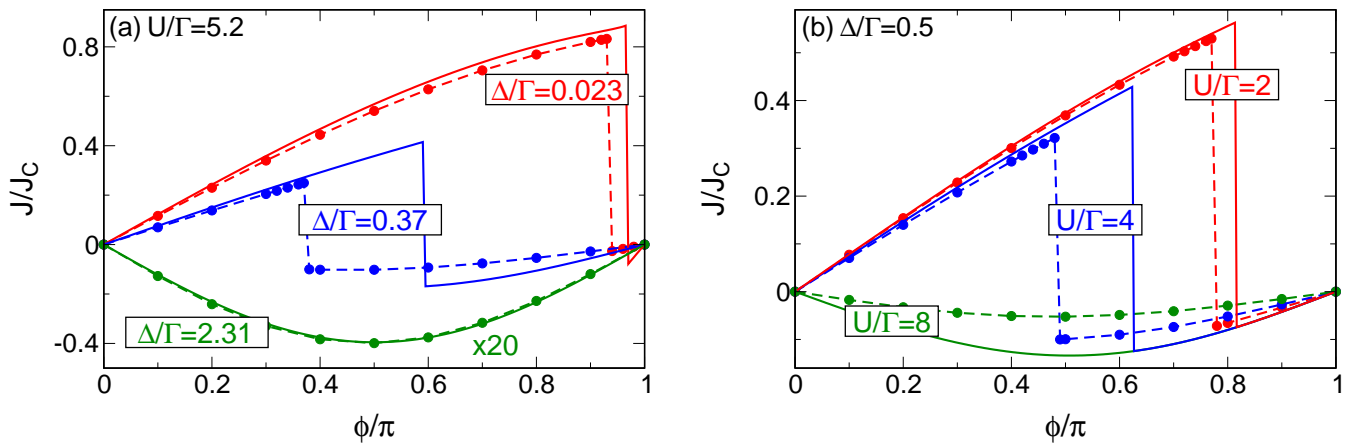

   \includegraphics[height=5.8cm,clip]{currentcomp2.eps}\hspace*{0.03\linewidth}
   \includegraphics[height=5.8cm,clip]{currentcomp1.eps}
   \caption{(Color online) Zero-temperature Josephson current $J$ in units of $J_c=e\Delta/\hbar$ as a function of the phase difference $\phi$ computed with fRG (solid lines) and NRG (symbols) at $\epsilon=0$ and $\Gamma_L=\Gamma_R$. (a) $\Delta$ is varied at fixed $U/\Gamma=5.2$ ($T_K/\Gamma=0.209)$. The parameters correspond to $\Delta/T_K=0.11$, $\Delta/T_K=1.76$, and $\Delta/T_K=11$ and were chosen for direct comparison with Fig.~3 of Ref.~\onlinecite{choi} (please note Ref.~\onlinecite{choiparameter}). For clarity, the curves at $\Delta/T_K=11$ were scaled up by a factor of $20$. (b) $\Delta/\Gamma=0.5$ is fixed at different $\Delta/T_K=1.1$, $\Delta/T_K=1.7$, and $\Delta/T_K=5.8$. The discretization parameter for NRG was chosen to be $\la=4$ (for details see Secs.~\ref{sec:methods.nrg}, \ref{sec:comp.noninteracting}, and Appendix \ref{sec:NRG_appendix}).}
   \label{fig:currentcomp}
\end{figure*}

\subsection{Phase diagrams}
\label{sec:results.phasediag}
Fig.~\ref{fig:phasediag1} shows the phase diagram at the point of particle-hole symmetry ($\epsilon=0$) for zero phase difference and symmetric couplings. It confirms that if either $U$ is increased or $\Gamma$ decreased, implying that the Bethe ansatz Kondo scale\cite{hewson}
\begin{equation}\label{eq:results.phasediag.TK}
T_K=\sqrt{U\Gamma/2}\exp\left[-\frac{\pi}{8U\Gamma}\left|U^2-4\epsilon^2\right|\right]
\end{equation}
decreases, the system shows a phase transition from a non-magnetic singlet to a magnetic doublet ground state.\cite{tkdelta} If $U$ and $\Gamma$ are fixed, the system is always in a singlet state at sufficiently small $\Delta$. As $\Delta$ becomes larger, a phase transition is observed provided that $U>2\Gamma$. In contrast, for $\Gamma>U/2$ the Kondo effect is not active and the ground state always remains a (BCS) singlet no matter how large $\Delta$. This can be understood in detail from the analytical treatment of the so-called atomic limit $\Delta=\infty$. The inset of Fig.~\ref{fig:phasediag1} illustrates that at large $\Delta$, the phase boundary indeed approaches the analytical result $\Gamma_c(U)=U/2$ [see Eq.~(\ref{eq:comp.atom.phaseboundary}) of Sec.~\ref{sec:comp.atom}].

A phase diagram for $\Gamma_L\neq\Gamma_R$ and finite phase difference $\phi$ is shown in Fig.~\ref{fig:phasediag2}. The general picture of the phase transition remains the same, only the phase boundary is affected. One can see that asymmetric couplings $\Gamma_L\neq\Gamma_R$ stabilize the singlet phase. In contrast, the doublet phase becomes more and more favorable with increasing $\phi$. The phase boundary continuously evolves from the $\phi=0$ into the $\phi=\pi$ curve if $\phi$ is gradually increased from $0$ to $\pi$. If $\Gamma_L=\Gamma_R$ and $\phi=\pi$, the singlet phase completely disappears (both within fRG and NRG). The phase boundary at finite $\epsilon$ (dotted lines) approximately starts from $U=2.5\Delta$, which is the point where in the atomic limit ($\Gamma/\Delta=0$) singlet and doublet states cross. Generalizing the arguments presented in Sec.~\ref{sec:comp.atom} for $\Gamma_L\neq\Gamma_R$,\cite{oguri2} it is possible to demonstrate that the boundary line shows square-root behaviour close to $\Gamma=0$. One should note that in our case the deviation of the dot energy from the point of particle-hole symmetry depends on the interaction strength. Hence, the Kondo temperature $T_K$ given by Eq.~(\ref{eq:results.phasediag.TK}) increases with $U$, causing the non-monotonicity in the phase boundary. Additional NRG results away from particle-hole symmetry can be found in Ref.~\onlinecite{oguri2}.

In general, the phase boundaries obtained from fRG and NRG agree quite well. As expected, the agreement is particularly good at small $U$. The dashed line in the main part of Fig.~\ref{fig:phasediag1} shows fRG results where the flow of the static vertex was discarded. Whereas the general physics is captured by this even simpler fRG truncation scheme (which sets $U^\la=U$), the quantitative agreement with NRG improves if the flow equation (\ref{eq:methods.frg.flowU}) is accounted for. Hence, we will stick to this improved approximation from now on.

\subsection{Josephson current}
\label{sec:results.current}
In Fig.~\ref{fig:currentcomp}, we show fRG and NRG results for the Josephson current $J$ as a function of the phase difference $\phi$. The current is given in units of $J_c=e\Delta/\hbar$. The figure illustrates the generic physical picture; in particular, we observe the same away from particle-hole symmetry ($\epsilon\neq0$) and for asymmetric couplings ($\Gamma_L\neq\Gamma_R$). In certain limits analytical results help to understand the form of $J(\phi)$. We will discuss this in detail in Sec.~\ref{sec:comp}. However, it is instructive to recall that the current flowing between two superconductors connected by a weak tunneling barrier is purely sinusoidal [$J\sim\sin(\phi)$].\cite{pinkbook}

As discussed above, the doublet phase is stabilized if $U$, $\Delta$, or $\phi$ is increased. Thus, at appropriate $\Delta/T_K$, the phase transition manifests as a discontinuity at $\phi_c$ in the $J(\phi)$ curve. This is illustrated in Fig.~\ref{fig:currentcomp}(a,b) (at fixed $T_K$ and $\Delta$, respectively). Note that the parameters of Fig.~\ref{fig:currentcomp}(a) were chosen such that we can directly compare our results with those of Ref.~\onlinecite{choi} (see Sec.~\ref{sec:comp.nrgqmc}).\cite{choiparameter} As there is a difference between the phase boundaries obtained from fRG and NRG, $\phi_c$ is different in both approaches. However, the form of the curves is captured well by fRG even at large $U/\Gamma$. The agreement between fRG and NRG concerning the current amplitude is good at small to intermediate $U$, being almost perfect throughout the singlet phase. This is satisfying as our truncated fRG is by construction well-controlled in this regime. The agreement becomes worse as $U$ becomes large [see Fig.~\ref{fig:currentcomp}(b)]. In general, the deviation between both methods is most severe in the doublet phase at intermediate $\Delta$, becoming better as $\Delta$ increases [compare Figs.~\ref{fig:currentcomp}(a) and (b)]. It is an artifact of the fRG approximation that the current is completely independent of $U$. However, the sinusoidal form of $J(\phi)$ in the doublet phase is described equally well by both methods.

\begin{figure}[tb]
   \includegraphics[height=5.8cm,clip]{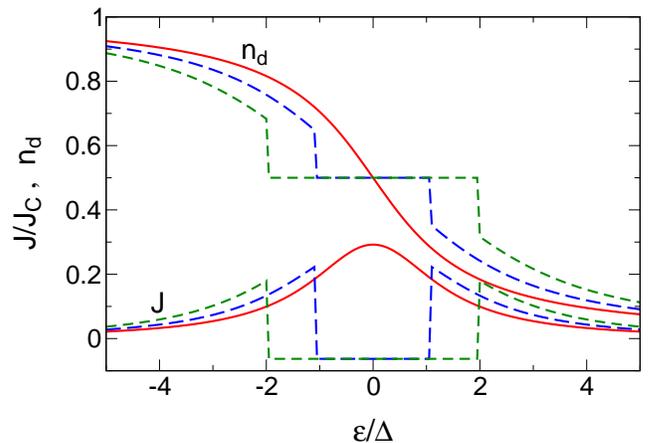}
   \caption{(Color online) fRG results for the Josephson current $J$ and the average occupation number $n$ at $T=0$ as a function of the dot's energy $\epsilon$ at $U/\Gamma=3$ (solid), $U/\Gamma=5$ (long dashed), and $U/\Gamma=7$ (short dashed). The other parameters are $\Delta/\Gamma=1$, $\phi/\pi=0.5$, and $\Gamma_L=\Gamma_R$. These results are very much consistent with recent experiments (see Fig.~4 of Ref.~\onlinecite{exprasmussen}).}
   \label{fig:current_epsilond}
\end{figure}

One parameter that can be easily tuned experimentally is the energy of the quantum dot,\cite{exprasmussen} rendering it reasonable to consider the current as a function of $\epsilon$. Within fRG, computing $J(\phi)$ away from half-filling is not different from computing $J(\phi)$ at $\epsilon=0$. In contrast, NRG greatly benefits from symmetry properties which only hold at $\epsilon=0$ (see Appendix \ref{sec:NRG_appendix}), and computing the current away from this point is (though possible) numerically demanding. As we have demonstrated fRG to provide acceptable accuracy in comparison with NRG at small to intermediate $U$, we refrain from this task and show only fRG results for $J(\epsilon)$ and the average occupation number $n(\epsilon)$ in Fig.~\ref{fig:current_epsilond}. As mentioned before, the singlet phase is always favored at large $|\epsilon|$. If $\Delta/T_K^{\epsilon=0}$ is appropriately chosen, the phase transition to the doublet ground state occurs at a certain $|\epsilon_c|$. The system is in the doublet phase for $|\epsilon|<|\epsilon_c|$, and both the current and the occupation $n_d(\epsilon)=1/2$ are independent of $U$ and $\epsilon$ within the fRG approximation. NRG calculations showed that the deviations from $n_d(\epsilon)=1/2$ are indeed small.\cite{oguri2} One should note that our findings for $J(\epsilon)$ are consistent with recent experiments.\cite{exprasmussen}

Within fRG, another possibility to compute the Josephson current is to differentiate the grand canonical potential $\Omega$ w.r.t.~the phase difference $\phi$ [see Eq.~(\ref{eq:currentconservation.currentbyenergy})]. Since the truncated fRG is a non-conserving approximation, the current computed in this way will generally differ from the one obtained via the self-energy formula Eq.~(\ref{eq:methods.frg.current}). The quantity $\Omega$ can be obtained directly from a flow equation. From the formalism one would expect the energy to be computed more accurately than the self-energy (the former being a vertex function of lower order), rendering it reasonable to calculate the Josephson current from the approximated energy rather than from the approximated self-energy. This expectation is contradicted by the observation that the former way does not give meaningful results in the doublet phase; in particular, the current remains positive. In the singlet phase the current computed from the energy compares badly to exact results obtained at $\Delta\to\infty$. Further studies on this issue are required. For the time being, we calculate $J$ from the self-energy via Eq.~(\ref{eq:methods.frg.current}).

\section{Other approaches}
\label{sec:comp}
In this section we discuss different approaches to the Josephson problem. An analytical treatment is possible if either $U=0$ or $\Delta=\infty$. Both cases were partly adressed in prior works (see e.g.~Refs.~\onlinecite{oguri2} and \onlinecite{vecino} for the atomic limit) but are again reported on here since the exactly-solvable noninteracting case is an important check for the NRG numerics, whereas analytical results at $U>0$ provide an additional benchmark for the fRG. We argue that a previous NRG approach by Choi et al.~(Ref.~\onlinecite{choi}) does not agree quantitatively with our results that we believe to be correct, although qualitatively both NRG approaches yield similar data except for a few obvious errors.\cite{choierror} We present NRG calculations of the Josephson current at finite temperatures and compare these results with the quantum Monte Carlo approach by Siano and Egger (Ref.~\onlinecite{se}). The discrepancy between both methods can be explained by considering the first excited many-body energies obtained from NRG. Finally, we comment on the issue of mean-field calculations.

\subsection{The noninteracting case}
\label{sec:comp.noninteracting}

\begin{figure}[tb]
   \includegraphics[height=5.87cm,clip]{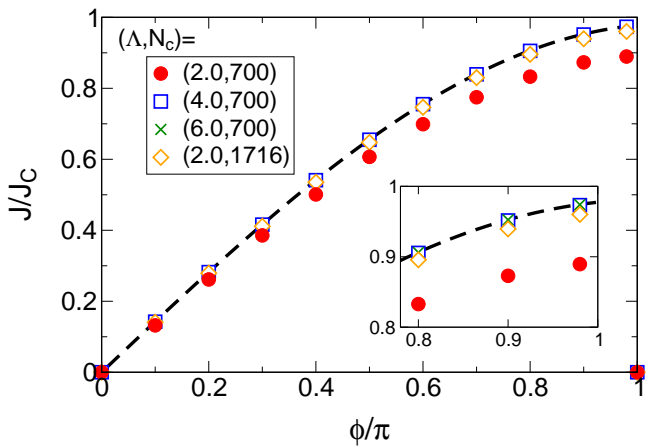}
   \caption{(Color online) NRG results for the zero-temperature Josephson current through a noninteracting, particle-hole symmetric dot for $\Delta/\Gamma=0.023076$ and several sets of the discretization parameter $\Lambda$ and the number of kept states $N_c$. The dashed line denotes the analytic result Eq.~(\ref{eq:currentnoninteracting3}). All other NRG results for $J$ in this paper were obtained at $\la=4$.}
   \label{fig:nrg}
\end{figure}

At $U=0$, the ground state of the system is always a (BCS) singlet. There is no doublet phase. An analytical expression for the zero-temperature Josephson current is provided by Eq.~(\ref{eq:methods.frg.current}). Setting the self-energy to zero (and focusing on $\Gamma_L=\Gamma_R$), the integral can be rewritten as
\begin{equation}\label{eq:currentnoninteracting3}
J = \frac{1}{2\pi}\int\frac{\Delta\sin(\phi)}{\cos^2(\phi/2)+x^2\left(1+(\Delta/\Gamma)y\right)^2+(\epsilon/\Gamma)^2y^2}\,dx,
\end{equation}
where $y=\sqrt{1+x^2}$. The integration can be carried out analytically if one expands in terms of $\Delta/\Gamma$, $\epsilon/\Gamma$ or $\Gamma/\Delta$, $\epsilon/\Delta$. One obtains
\begin{equation}\label{eq:currentnoninteracting2}
\lim_{ \left\{\Delta\to\infty \atop \Gamma\to\infty\right\} } J
= \left\{ \Gamma \atop \Delta \right\} \frac{\sin(\phi)}{2\sqrt{(\epsilon/\Gamma)^2+\cos^2(\phi/2)}}.
\end{equation}
This shows that $J$ is of order $\Gamma$ ($\Delta$) at large $\Delta$ ($\Gamma$). A small $\epsilon\neq 0$ affects the current mainly for $\phi\gtrsim2\arccos(\epsilon/\Gamma)$. Running fRG, we reproduce these results. This is, however, only a consistency check for our numerics but not for the fRG approximation (which by construction is exact at $U=0$) itself. In contrast, reproducing noninteracting results is an important check for the NRG. Particularly for two-channel models, one cannot expect such a high numerical accuracy as one achieved for single-channel models. We have checked our NRG data in several ways. Fig.~\ref{fig:nrg} shows an example, which is the Josephson current at $U=0$ in the particle-hole symmetric case. Here, $\Gamma$ is chosen to be much larger than the gap, $\Gamma=43.2\Delta$. The NRG results are examined for several sets of ($\Lambda$, $N_c$) given by ($\square$: 4.0, 700), ($\bullet$: 2.0, 700), ($\diamond$: 2.0, 1716), and ($\bm{\times}$: 6.0, 700). The NRG data can be compared with the exact $U=0$ result (dashed line). As we discussed in Sec.~\ref{sec:methods.nrg}, the numerical accuracy is controlled by the discretization parameter $\Lambda$ and the number of the low-energy states $N_c$ retained in the truncation procedure for constructing the Hilbert space for the next NRG step. The results show that for $\Lambda \gtrsim 4$, the NRG iterations with $700$ kept states generate correct convergent data (see the points corresponding to $\square$ and $\bm{\times}$ in the inset). On the other hand, the data for a small discretization $\Lambda=2$ shows that $700$ states are not enough, and the results ($\bullet$) deviate from the exact ones. However, with a large number of the kept states ($N_c=1716$), the numerical accuracy can be improved nicely, as the results plotted with $\diamond$ approach to the correct ones. The dependence of the convergence on the truncation procedure becomes sensitive particularly at $\Gamma \gg \Delta$, when the wave function of the Andreev bound state penetrates deep into the superconducting leads. We have confirmed that all statements from above also found for finite $U$. From these checks, we see that for the two-channel model the truncation has to be performed with special care, in particular for a small discretization such as $\Lambda=2$. We have also confirmed that for larger discretizations $\Lambda \simeq 4$, $700$ states are enough to obtain convergent results in the interacting case. Therefore, in this paper we have chosen $\Lambda=4$ for all other NRG results for $J$, and have benchmarked the convergence at some data points on each of the curves by carrying out cross-checks using iterations with $1716$ kept states.

\begin{figure*}[tb]
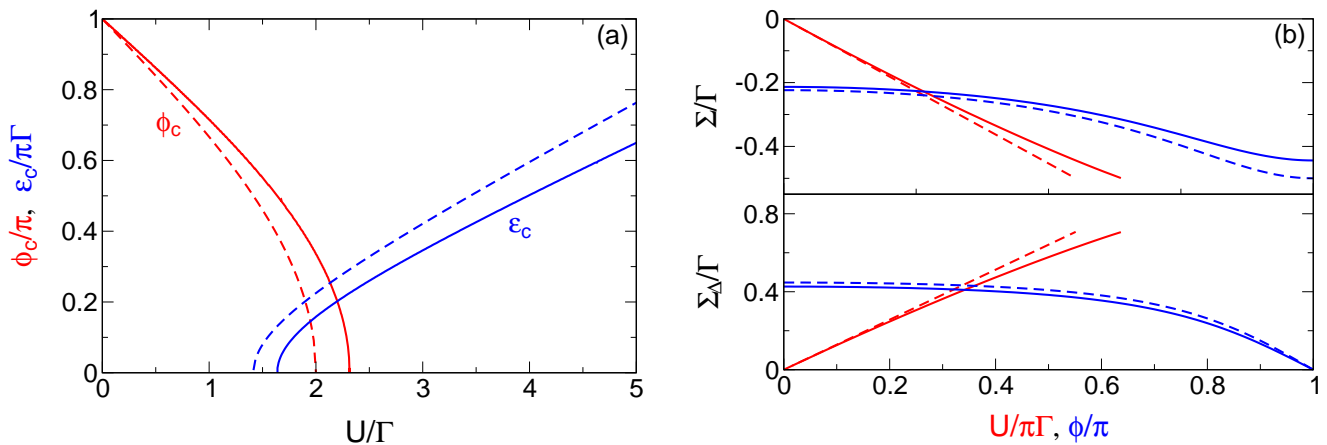

   \includegraphics[height=5.8cm,clip]{atom1.eps}\hspace*{0.03\linewidth}
   \includegraphics[height=5.8cm,clip]{atom2.eps}
   \caption{(Color online) (a) Comparison of fRG (solid) with the exact result [dashed, from Eq.~(\ref{eq:comp.atom.phaseboundary})] for the critical phase difference $\phi_c$ at $\epsilon=0$ and the critical gate voltage $\epsilon_c$ at $\phi/\pi=0.5$ describing the phase boundary in the atomic limit. fRG results were obtained at $\Delta/\Gamma=1000$. (b) The same, but comparing the $U$ - and $\phi$ - dependence (at $\epsilon/\Gamma=0.5$, $\phi/\pi=0.5$ and $\epsilon/\Gamma=0.5$, $U/\Gamma=1$, respectively) of the self-energy components. Only the results in the singlet phase are shown. The end of the lines $\Sigma(U)$ and $\Sigma_\Delta(U)$ indicate the transition into the doublet phase.  }
   \label{fig:atom1}
\end{figure*}

\subsection{The atomic limit}
\label{sec:comp.atom}
It is instructive to investigate the so-called atomic limit $\Delta=\infty$ at finite $U>0$. Even though the current in the doublet phase vanishes at $\Delta=\infty$, one can obtain analytical expressions for $J$ in the singlet phase as well as for the phase boundary. Here, we focus on symmetric couplings, the more general situation with $\Gamma_L\neq\Gamma_R$ is discussed in Ref.~\onlinecite{oguri2}.

At $\Delta=\infty$, the inverse of the free propagator (\ref{eq:methods.frg.g0}) reduces to
\begin{equation}
\left[\mc G^0(i\omega)\right]^{-1} = 
\begin{pmatrix}
i\omega - \epsilon & \tilde\Delta_d \\ \tilde\Delta_d & i\omega + \epsilon
\end{pmatrix},
\end{equation}
where $\tilde\Delta_d=\Gamma\cos(\phi/2)$. Including the interacting part, the problem becomes equivalent to solving an effective two-level Hamiltonian
\begin{equation}\begin{split}
H_\tn{eff} = ~& \epsilon\left(\varphi_1^\dagger\varphi_1-\varphi_2^\dagger\varphi_2\right)
-\tilde\Delta_d\left(\varphi_1^\dagger\varphi_2+\varphi_2^\dagger\varphi_1\right) \\
& -U\left(\varphi_1^\dagger\varphi_1-\frac{1}{2}\right)\left(\varphi_2^\dagger\varphi_2-\frac{1}{2}\right).
\end{split}\end{equation}
Diagonalizing this operator yields a non-degenerate [doubly degenerate] ground state if $B(\phi)>U/2$ [$B(\phi)<U/2$], where $B(\phi)=\sqrt{\epsilon^2+\tilde\Delta_d^2}$. Hence, the boundary between the singlet and doublet phase is described by
\begin{equation}\label{eq:comp.atom.phaseboundary}
\frac{1}{4}\left(\frac{U}{\Gamma}\right)^2-\left(\frac{\epsilon}{\Gamma}\right)^2-\cos^2(\phi/2)=0.
\end{equation}
This shows that at small $U$ the system never exhibits a phase transition no matter how large $\Delta$. The Kondo effect is not active and the ground state always remains a (BCS) singlet. As discussed above, this singlet phase stabilizes with increasing $\epsilon$ and decreasing $\phi$. A non-trivial test for the fRG approximation is to compare critical lines obtained numerically at large $\Delta/\Gamma$ with the exact result Eq.~(\ref{eq:comp.atom.phaseboundary}). This is done in Fig.~\ref{fig:atom1}(a), showing that fRG captures the essential behavior of the critical lines.

\begin{figure}[tb]
   \includegraphics[height=5.8cm,clip]{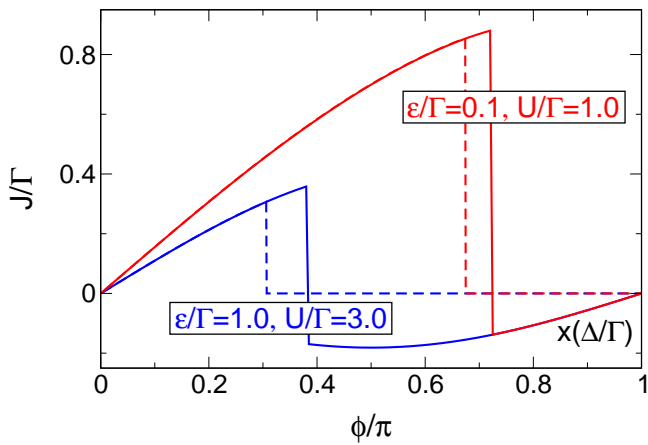}
   \caption{(Color online) Josephson current obtained from fRG (solid lines) in the atomic limit $\Delta=\infty$ at $T=0$. As fRG calculations were performed at $\Delta/\Gamma=1000<\infty$, the current in the doublet phase is finite due to $\Gamma/\Delta$ corrections [see Eq.~(\ref{eq:results.flow.current})]. Note that in this phase $J$ (which was scaled up by a factor of $\Delta/\Gamma$) is independent of $U$ and $\epsilon$. The dashed lines (which are partly covered by the solid lines on this scale) show the analytical result for the singlet phase [Eq.~(\ref{eq:comp.atom.current})]. }
   \label{fig:atom2}
\end{figure}

The Josephson current in the singlet phase follows from differentiating the energy of the corresponding state $E=U/4-B(\phi)$ w.r.t.~the phase difference $\phi$ [see Eq.~(\ref{eq:currentconservation.currentbyenergy})]:
\begin{equation}\label{eq:comp.atom.current}
J = \frac{\Gamma}{2}\frac{\sin(\phi)}{\sqrt{(\epsilon/\Gamma)^2+\cos^2(\phi/2)}}.
\end{equation}
This coincides with the noninteracting result Eq.~(\ref{eq:currentnoninteracting2}). Since the energy of the doublet state $E=-U/4$ is independent of $\phi$, the current in the doublet phase vanishes. In Fig.~\ref{fig:atom2} we show that the current (in the singlet phase) obtained from fRG at large $\Delta/\Gamma$ indeed falls onto the curve described by Eq.~(\ref{eq:comp.atom.current}).

The knowledge of the exact eigenstates also allows us to compute the exact Green function (using the spectral representation) and from this the exact self-energy.\cite{oguri2} In the singlet phase the components are given by
\begin{equation}
\Sigma(i\omega) = -\frac{U\epsilon}{2B(\phi)}~~~~~~~~\Sigma_\Delta(i\omega)=\frac{U\tilde\Delta_d}{2B(\phi)}.
\end{equation}
The self-energy is purely of first order in $U$ and frequency-independent. In Fig.~\ref{fig:atom1}(b) we compare the fRG approximated self-energy to this exact result. Overall the fRG reproduces the parameter dependencies quite well. However, compared to the exact result the fRG self-energy contains higher-order corrections which for larger $U/\Gamma$ lead to deviations from the strictly linear $U$ - dependence. In the doublet phase the exact self-energy is given by
\begin{equation}
\Sigma(i\omega) =-\frac{U^2}{4}\frac{i\omega+\epsilon}{\omega^2+[B(\phi)]^2}~~~~~
\Sigma_\Delta(i\omega) = \frac{U^2}{4}\frac{\tilde\Delta_d}{\omega^2+[B(\phi)]^2}.
\end{equation}
Remarkably it is purely of second order in $U$ and in contrast to the self-energy in the singlet phase frequency-dependent. One cannot expect these properties to be reproduced by the truncated fRG which neither keeps all terms of order $U^2$ nor any frequency dependence. Indeed, the fRG self-energy in the doublet phase is always (i.e.~also in the atomic limit) given by $\Sigma=-\epsilon$ and $\Sigma_\Delta=\tilde\Delta_d$.

To conclude, we have demonstrated fRG to well reproduce analytical results for $\Delta=\infty$. As mentioned in Sec.~\ref{sec:methods.nrg}, NRG is not as accurate for two-channel models as it is for single-channel models, and in our case only the first several digits of the expectation values can be trusted. Thus, for calculating the very small current $J/J_c$ (of order $\lesssim 10^{-3}$)  accurately in a doublet state (particularly at $\Gamma \ll \Delta$ for large $U$), it is still not efficient to use NRG at present.

\subsection{NRG vs.~QMC}
\label{sec:comp.nrgqmc}
The Josephson current was previously computed using NRG by Choi et al.~(CLKB, Ref.~\onlinecite{choi}). The accuracy of their results was questioned by Siano and Egger (SE, Ref.~\onlinecite{se}) who performed quantum Monte Carlo (QMC) calculations. However, this issue has not been finally resolved yet.\cite{disputechoise1,disputechoise2} Furthermore, QMC is an inherently finite-$T$ method, whereas CLKB's NRG approach mainly focused on the zero temperature limit. Since both methods are generally regarded to benefit from numerical exactness, more clarifying work is needed.

\begin{figure}[tb]
   \includegraphics[height=5.8cm,clip]{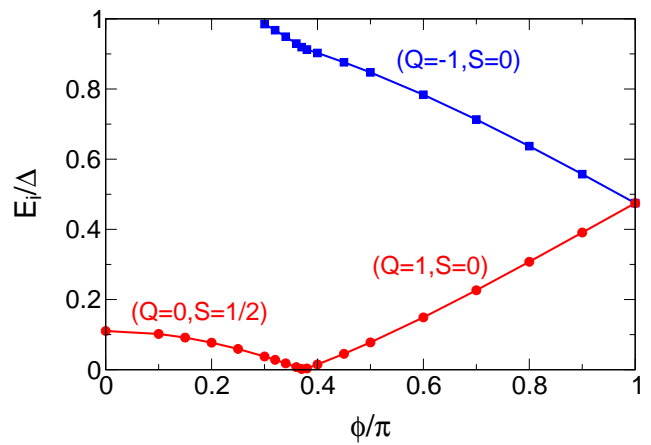}
   \caption{(Color online) Energies $E_i$ of the first and second many-body excited state emerging below the edge of the superconducting gap at $U/\Gamma=5.2$, and $\Delta/\Gamma=0.37$. $E_i$ is measured with respect to the ground-state energy. $(Q,S)$ denotes the set of quantum numbers introduced in Sec.~\ref{sec:methods.nrg}. The NRG parameters are taken to be same as in Fig.~\ref{fig:currentcomp}(a).}
   \label{fig:nrgE}
\end{figure}

In the present paper we have re-examined the NRG calculation for the Josephson current checking the numerical accuracy carefully as discussed in Sec.~\ref{sec:comp.noninteracting}. The parameters in Fig.~\ref{fig:currentcomp}(a) are chosen for direct comparison to Fig.~3 of CLKB's paper.\cite{choiparameter,choilambda} Whereas the data we obtained from fRG and NRG agree fairly well, there is a sizable discrepancy to CLKB's results. Most important, the amplitude of $J$ exceeds the one computed by CLKB by a factor of about 1.5 - 2. Our results suggest that CLKB's data for the Josephson current have a problem in the amplitude, although their results capture the $\phi$ dependence correctly except in the region near $\phi = \pi$.\cite{choierror} In addition, we have studied the $\phi$ - dependence of the energies $E_i$ of the first and second many-body excited states (see Fig.~\ref{fig:nrgE}), which emerge below the gap ($0<E_i<\Delta$). These in-gap excitations determine the temperature dependence at $T\ll\Delta$ (see below). We observe that our results of the first excited energy agree reasonably well with the peak position of the single-particle spectrum given in Fig.~2(d) of Ref.~\onlinecite{choi}, although the broadening of the bound-state peak seen in the figure of CLKB must be artificial. One should note that the first and second excited states are degenerate at $\phi=\pi$. This is caused by a special symmetry holding at $\phi=\pi$ between eigenstates with the quantum number $Q$ and those with $-Q$.\cite{oguri2} 

\begin{figure}[tb]
   \includegraphics[height=5.8cm,clip]{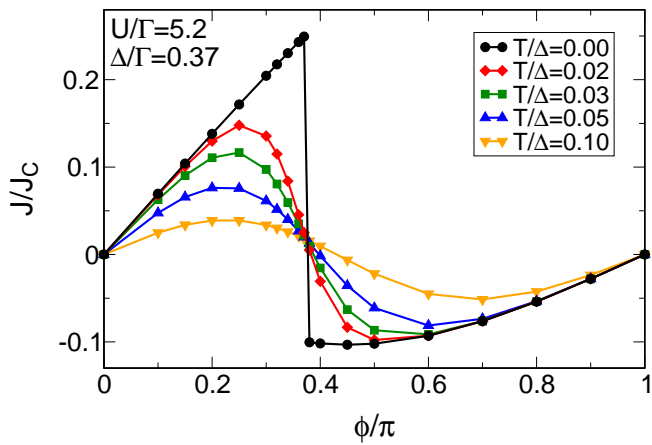}
   \caption{(Color online) NRG results for the Josephson current at finite temperatures $T$. The parameters are the same as in Fig.~\ref{fig:nrgE}.}
   \label{fig:nrgT}
\end{figure}

In order to further clarify the discrepancies between the previous NRG and QMC results, we have re-examined the temperature dependence of the Josephson current using our NRG code. The results are shown in Fig.~\ref{fig:nrgT}, where the parameters are taken to be the same as those used for the $T=0$ results at $\Delta/\Gamma = 0.37$ given in Fig.~\ref{fig:currentcomp}(a). Thus, our data can be directly compared with Fig.~1 of Ref.~\onlinecite{disputechoise1}. The general features of the $T$ - dependence of CLKB's results are consistent with our findings, although the amplitudes again differ approximately by a factor of the order of 2.

Figure \ref{fig:nrgT} can also be compared with SE's QMC results, namely with Fig.~1 of Ref.~\onlinecite{disputechoise1}. We observe that the amplitude of the Josephson current obtained by SE is consistent with ours in the region $\pi/2\lesssim\phi<\pi$ where the ground state is a doublet. However, SE's results of the current are very small at $0\leq\phi\lesssim 0.2\pi$ compared to our NRG data. The reason for the discrepancy may be inferred from the $\phi$ - dependence of the excitation energy. As shown in Fig.~\ref{fig:nrgE}, the first excitation energy $E_1$ is very small ($0<E_1\lesssim 0.1\Delta$) for $0\leq\phi\lesssim \pi/2$. In particular, $E_1$ is smaller than (or almost equal to) the temperature $T=0.1\Delta$ that SE used for their calculations. They have set up the transition probability for the importance sampling introducing the ultraviolet cutoff for the summations over the Matsubara frequencies such that $|\omega_n|<\pi PT$, where $P$ is the Trotter number.\cite{se,QMC_normal} Therefore it is questionable whether the transition probability captures accurately the information about the in-gap states if the excitation energies are very small. This will not cause any problems if $E_1$ is larger than $T$, and it explains naturally the agreement between the QMC and our NRG data for $\pi/2\lesssim\phi\leq\pi$.

\subsection{Mean-field}
\label{sec:comp.hf}

\begin{figure}[tb]
   \includegraphics[height=5.8cm,clip]{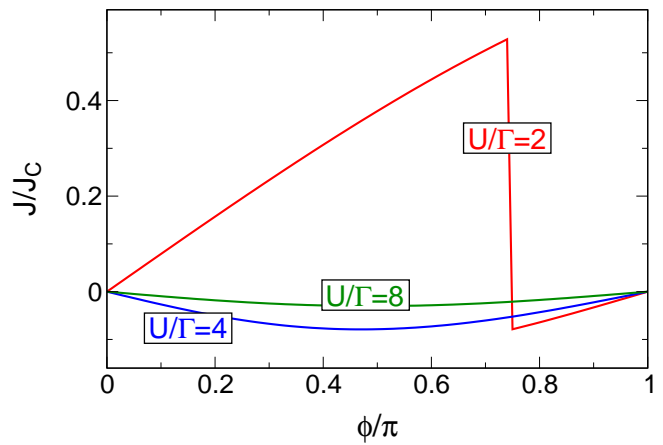}
   \caption{(Color online) The same as in Fig.~\ref{fig:currentcomp}(b), but obtained within a self-consistent Hartree-Fock framework. }
   \label{fig:hf}
\end{figure}

For the ordinary single impurity Anderson model it is well-known that the Kondo effect cannot be described within a mean-field framework. Despite this fact the Hartree-Fock approximation was used to compute the Josephson current through a single impurity coupled to BCS leads. However, previous approaches were either incomplete (the induced anomalous term was discarded in Ref.~\onlinecite{ra}) or inaccurate (Ref.~\onlinecite{yo}; see below). For reference, we have thus performed our own Hartree-Fock calculation based on self-consistent equations which we derived in agreement with Yoshioka and Ohashi (YO, Ref.~\onlinecite{yo}). As within truncated fRG, the Hartree-Fock self-energy is frequency-independent. Astonishingly, the general picture of the phase transition is captured on this mean-field level, but the quantitative agreement with reliable NRG data is poor [compare Figs.~\ref{fig:currentcomp}(b) and \ref{fig:hf}]. However, the observation that Hartree-Fock succeeds in qualitatively describing the behaviour of the Josephson current is an uncontrolled result since an approximation which inherently does not contain the Kondo temperature cannot be expected to describe a transition governed by $\Delta/T_K$. In addition, the appearance of a phase with $J<0$ is caused by an unphysical breaking of the spin symmetry (the ground state is not a doublet). Thus, a direct comparison between Hartree-Fock and NRG is actually of limited value and provided by Fig.~\ref{fig:hf} only for reasons of completeness.

It is important to mention that even though we derive the same mean-field equations as YO, we cannot reproduce their numerical solution. In particular, YO observe the induced magnetization (the difference between the average occupation numbers of spin up and down electrons) to increase continuously from zero when $U$ is increased. In contrast, our solution shows a discontinuous jump when the system enters the phase with $J<0$. We have double-checked our data using different routines to solve the self-consistency problem and are thus tempted to conclude that YO's results are inaccurate due to numerical issues.

\section{Conclusions}
\label{sec:conclusions}
In this paper we have presented phase diagrams and the Josephson current for a single Anderson impurity coupled to BCS superconducting leads. We obtained our data using the frameworks of the functional and numerical renormalization group, respectively. The well-known accuracy of the NRG was established by comparisons with the exact solution at $U=0$, allowing us to show that previous NRG results for the Josephson current by Choi et al.~are smaller approximately by a factor $2$ compared to the correct values. Using NRG we have also re-examined the temperature dependence of the current. Our results agree reasonably well with the QMC data by Siano and Egger when the in-gap excited energy $E$ is larger than $T$, while in the opposite case $0\leq E\lesssim T < \Delta$ the QMC results become less accurate. We have demonstrated the truncated fRG to perform well compared with NRG at small to intermediate interactions and to reproduce analytical results which we derived in the limit $\Delta=\infty$. As fRG requires virtually no numerical resources it can be used to fast provide current - phase curves needed to extract physical quantities out of experimental data.\cite{expnovotny} Concerning this issue it would be desirable to extend the fRG scheme to treat finite temperatures. Doing this in a reliable but at the same time numerically efficient way is difficult and needs further investigation.

\section*{Acknowledgments}
This work was initiated by enlightening discussions with T.~Novotn\'y and J.~Paaske when V.M.~was visiting the Nano-Science Center of the University of Copenhagen. V.M.~very much enjoyed the hospitality of this institution. A.O.~is grateful to the condensed matter physics group of the Universit\"at G\"ottingen for the kind hospitality during his stay. His visit was supported by the Deutsche Forschungsgemeinschaft (SFB 602). We are much obliged to S.~Andergassen, J.~Bauer, A.~C.~Hewson, Y.~Nisikawa, T.~Novotn\'y, T.~Pruschke, K.~Sch\"onhammer, Yoichi Tanaka, and Yoshihide Tanaka for valuable discussions and thank P.~Pl\"otz for careful reading of the manuscript. C.K.~and V.M.~are grateful to the Deutsche Forschungsgemeinschaft (FOR 723) for support. A.O.~is supported by the Grant-in-Aid for Scientific Research from JSPS. NRG calculations were partly carried out on SX8 at the computation center of the Yukawa Institute.

\appendix
\section{NRG approach}
\label{sec:NRG_appendix}
In the NRG approach, the continuous conduction bands of the original model Hamiltonian are discretized and described by the fermions $f_{L,n\sigma}^{\phantom{\dagger}}$ and $f_{R,n\sigma}^{\phantom{\dagger}}$ defined on a linear chain for $n\geq 0$ on the left and right, respectively.\cite{nrg1} In the case of equal gaps ($\Delta_L=\Delta_R=\Delta$) and symmetric couplings ($\Gamma_L=\Gamma_R=\Gamma/2$) it is convenient to use the linear combinations\cite{choi}
\begin{equation}\begin{split}
s_{n\sigma}&=\frac{e^{-i\phi/4}f_{R,n\sigma}+e^{i\phi/4}f_{L,n\sigma}}{\sqrt{2}},\\
p_{n\sigma}&=\frac{e^{-i\phi/4}f_{R,n\sigma}-e^{i\phi/4}f_{L,n\sigma}}{-\sqrt{2}i}.
\end{split}\end{equation}
The discretized Hamiltonian of the NRG,
\begin{equation}\label{eq:H_NRG}
H_\tn{NRG}=H^\tn{dot}+H^\tn{lead}_\tn{NRG}+H^\tn{coup}_\tn{NRG},
\end{equation}
can be expressed in a way such that it contains only real parameters for arbitrary phase difference $\phi$, even though the operator $H$ defined in Sec.~\ref{sec:model} is generally a complex Hermitian matrix when $\phi$ is finite. Namely, the part of the Hamiltonian modeling the BCS leads can be written as
\begin{equation}\label{eq:H_nrg_lead}\begin{split}
H_\tn{NRG}^\tn{lead} = \sum_{n=0}^\infty\Bigg[
 \sum_\sigma\,u_n&\Big(s_{n+1\sigma}^\dagger s_{n\sigma}+p_{n+1\sigma}^\dagger p_{n\sigma} +\tn{H.c.}\Big) \\
 + \,\Delta&\Big(s_{n\uparrow}^\dagger s_{n\downarrow}^\dagger-p_{n\uparrow}^\dagger p_{n\downarrow}^\dagger+\tn{H.c.}\Big)\Bigg].
\end{split}\end{equation}
The hopping matrix elements fall off exponentially with the distance $n$ from the dot:
\begin{equation}
u_n=D\frac{1+1/\la}{2}\frac{(1-1/\la^{n+1})\la^{-n/2}}{\sqrt{1-1/\la^{2n+1}}\sqrt{1-1/\la^{2n+3}}}.
\end{equation}
The discretized version of the coupling Hamiltonian reads
\begin{equation}\label{eq:H_mix_sp}\begin{split}
H_\tn{NRG}^\tn{coup} ~=~ & \sqrt{2}v_\tn{NRG}\cos(\phi/4)
\sum_\sigma\left(s_{0\sigma}^\dagger d_\sigma+d_\sigma^\dagger s_{0\sigma}\right) \\
+&\sqrt{2}v_\tn{NRG}\sin(\phi/4)
\sum_\sigma\left(p_{0\sigma}^\dagger d_\sigma+d_\sigma^\dagger p_{0\sigma}\right),
\end{split}\end{equation}
where we have defined $v_\tn{NRG}=\sqrt{2D\Gamma_\tn{NRG}/\pi}$, and
\begin{equation}\label{eq:A_L}
\Gamma_\tn{NRG}=\frac{A_\la\Gamma}{2}~~,~~~~ A_\la=\frac{1}{2}\frac{1+1/\la}{1-1/\la}\ln\la.
\end{equation}
The factor $A_{\Lambda}$ is necessary for correctly reproducing the original model in the continuum limit $\Lambda \to 1$.\cite{nrg1,SakaiShimizuKasuya,thomasnrg}
The discretized version of the current operator corresponds to a derivative of the tunneling Hamiltonian w.r.t.~the phase difference $\phi$:
\begin{equation}\label{eq:J_sp}
J_\tn{NRG}=2\partial_\phi H^\tn{coup}_\tn{NRG}.
\end{equation}
For a numerical implementation it is useful to exploit symmetry properties in order to reduce the size of the matrices to be diagonalized in each NRG step. Particularly, the pseudo-spin $Q^\tn{ps}$ defined by\cite{oguri2}
\begin{equation}\label{eq:I_x^sp}
Q^\tn{ps} = d_\uparrow^\dagger d_\downarrow^\dagger + \sum_{n=0}^\infty (-1)^{n+1}\left(s_{n\uparrow}^\dagger s_{n\downarrow}^\dagger+p_{n\uparrow}^\dagger p_{n\downarrow}^\dagger\right) +\tn{H.c.}
\end{equation}
is conserved, $[Q^\tn{ps},H_\tn{NRG}]=0$, in the particle-hole symmetric case ($\epsilon=0$). Therefore, the eigenvalue $Q$ of the operator $Q^\tn{ps}$ can be used as a quantum number to classify the Hilbert space in addition to the total spin $S$ associated with the rotational symmetry of the real spin.\cite{SSSS,SSSS_2} 

\section{Current formula}
\label{sec:currentformula}
 In this section we derive an exact formula relating the Josephson current to the self-energy. To this end, we define a current operator at lead $s=L,R$ as usual,
\begin{equation}
J_s=\partial_t N_s = i[H,N_s].
\end{equation}
Two terms of the Hamiltonian fail to commute with the particle number operator $N_s$:
\begin{equation}\label{eq:currentformula.currentop}\begin{split}
& \left[H_s^\tn{coup}+H_s^\tn{lead},N_s\right] \\[1ex]
&\hspace*{0.5cm}= t_s\sum_\sigma c_{s\sigma}^\dagger d_\sigma + 2\Delta_se^{i\phi_s}\sum_kc_{sk\uparrow}^\dagger c_{s-k\downarrow}^\dagger -\tn{H.c.}~.
\end{split}\end{equation}
The expectation value of the second term vanishes since $\Delta_se^{i\phi_s}\sim\sum_k\langle c_{sk\uparrow}c_{s-k\downarrow}\rangle$. The first term can be evaluated using a projection technique for the noninteracting Green function at the dot-lead interface:\cite{taylor}
\begin{equation}\label{eq:currentformula.projection1}
\mc G^0_{i,j}(z) = -t_s\left[\mc G^0(z)\tau_3g^s(z)\right]_{ij},
\end{equation}
with $\{i,j\}=1,2$ denoting Nambu indices of the dot and of the local site at the end of lead $s$, respectively. The generalization of Eq.~(\ref{eq:currentformula.projection1}) to the interacting Green function is achieved straight-forwardly by virtue of the Dyson equation:
\begin{equation}\label{eq:currentformula.projection2}
\mc G(z) = \mc G^0(z)+\mc G(z)\Sigma(z)\mc G^0(z).
\end{equation}
Using Eqs.~(\ref{eq:currentformula.projection1}) and (\ref{eq:currentformula.projection2}) to evaluate the expectation value of Eq.~(\ref{eq:currentformula.currentop}) we obtain the following expression for the current (see also Ref.~\onlinecite{oguri2}):
\begin{equation}\label{eq:currentformula.current1}
\langle J_s\rangle =
\frac{4t_s^2}{\beta}\sum_{i\omega}\tn{Im}\left[\mc G_{2,1}(i\omega)g^s_{1,2}(i\omega)\right],
\end{equation}
with $\mc G(z)$ being the dot Green function, and $\beta=1/T$. The most general form (in absence of a magnetic field) of $\mc G(i\omega)$ is
\begin{equation}\label{eq:currentformula.g}
\mc G(i\omega) = -\frac{1}{D(i\omega)}
\begin{pmatrix}
i\tilde\omega + \epsilon + \Sigma(i\omega)^* & \Sigma_\Delta(i\omega) -\tilde\Delta \\
\Sigma_\Delta(i\omega)^* -\tilde\Delta^* & i\tilde\omega - \epsilon - \Sigma(i\omega)
\end{pmatrix},
\end{equation}
where $D(i\omega)$ denotes the determinant
\begin{equation}\begin{split}
D(i\omega) = ~\,& \Big[\tilde\Delta(i\omega)-\Sigma_\Delta(i\omega)\Big]\Big[\tilde\Delta(i\omega)^*-\Sigma_\Delta(i\omega)^*\Big] \\
- & \Big[i\tilde\omega-\epsilon-\Sigma(i\omega)\Big]\Big[i\tilde\omega+\epsilon+\Sigma(i\omega)^*\Big].
\end{split}\end{equation}
The self-energy components fulfill the symmetry relations $\Sigma(-i\omega)=\Sigma(i\omega)^*$ and $\Sigma_\Delta(-i\omega)=\Sigma_\Delta(i\omega)$,\cite{oguri3} implying that $D(i\omega)$ is purely real. Employing Eqs.~(\ref{eq:currentformula.current1}) and (\ref{eq:currentformula.g}) allows for recasting the current formula into the simple form
\begin{equation}\label{eq:currentformula.current2}\begin{split}
\langle J_s\rangle = \frac{4}{\beta}\sum_{i\omega}\Bigg\{ &\frac{\Gamma_s\Delta_s\Gamma_{\bar s}\Delta_{\bar s}}{\sqrt{\omega^2+\Delta_s^2}\sqrt{\omega^2+\Delta_{\bar s}^2}} \frac{\sin(\phi_s-\phi_{\bar s})}{D(i\omega)} \\[1ex]
&+\frac{\Gamma_s\Delta_s\,\tn{Im}\left[e^{-i\phi_s}\Sigma_\Delta(i\omega)\right]}{D(i\omega)\sqrt{\omega^2+\Delta_s^2}}\Bigg\},
\end{split}\end{equation}
where we have introduced the notation $\bar{L}=R, \bar{R}=L$. One should note that this is an exact result, the generalization to the case of broken spin symmetry (i.e.~in presence of a magnetic field) being achieved straight-forwardly. At zero temperature, the Matsubara sum can be evaluated as an integral by replacing $1/\beta\sum_{i\omega}\rightarrow1/2\pi\int d\omega$. Within fRG and for symmetric parameters, $\Sigma_\Delta$ is real and we obtain Eq.~(\ref{eq:methods.frg.current}).

\section{Current conservation}
\label{sec:currentconservation}
This section is devoted to the question of current conservation, an issue which can also be tackled within a more general framework using generating functionals.\cite{oguri4} Within the model Hamiltonian Eq.~(\ref{eq:model.h}), electrons cannot be created or annihilated on the quantum dot. Hence, one would expect the Josephson current to be conserved:
\begin{equation}\label{eq:currentconservation.conservation}
\langle J_L\rangle = - \langle J_R\rangle .
\end{equation}
This can indeed be shown analytically by applying a gauge transformation $c_{sk\sigma}\to e^{-i\phi_s/2}c_{sk\sigma}$, $d_\sigma\to e^{-i\phi_R/2} d_\sigma$, and $H\to\tilde H$, with 
\begin{equation}
\tilde H(\phi) = H (\phi_L=\phi_R=0,t_L\to e^{-i\phi/2}t_L).
\end{equation}
The current operator Eq.~(\ref{eq:currentformula.currentop}) can then be expressed as a derivative of the grand canonical potential $\Omega$ w.r.t.\ the phase difference $\phi=\phi_L-\phi_R$:
\begin{equation}\label{eq:currentconservation.currentbyenergy}
\langle J_L\rangle = 2\langle\partial_\phi \tilde H(\phi)\rangle = 2\partial_\phi\Omega(\phi).
\end{equation}
The same result is obtained for $-\langle J_R\rangle$. Thus, the Josephson current is conserved.

Current conservation implies a symmetry relation for the self-energy. In particular, plugging Eq.~(\ref{eq:currentformula.current2}) into Eq.~(\ref{eq:currentconservation.conservation}) yields
\begin{equation}\label{eq:currentconservation.se}
\tn{Im}\sum_{s=L,R}\sum_{i\omega} \frac{\tilde\Delta_s(i\omega)\Sigma_\Delta(i\omega)^*}{D(i\omega)} =0.
\end{equation}
This equation is fulfilled if the interacting many-particle system is solved exactly. On the other hand, any approximate method to calculate the self-energy is current-conserving if and only if Eq.~(\ref{eq:currentconservation.se}) holds.

To discuss the issue of current conservation for frequency-independent approximations (such as truncated fRG and Hartree-Fock), it is instructive to rewrite the off-diagonal component of the self-energy in terms of a function $g(i\omega)=g(-i\omega)$ defined by
\begin{equation}
\Sigma_\Delta(i\omega) = g(i\omega)\sum_s\sum_{i\nu}a_s(i\nu),
\end{equation}
where $a_s(i\omega)=\tilde\Delta_s(i\omega)/D(i\omega)$. Equation (\ref{eq:currentconservation.se}) can then be recast as
\begin{equation}
\tn{Im}\,\sum_{s_1,s_2}\sum_{i\omega,i\nu}a_{s_1}(i\omega)a_{s_2}^*(i\nu)g^*(i\omega)=0.
\end{equation}
It follows that for any frequency-independent approximation to the self-energy, current conservation is equivalent to
\begin{equation}
\tn{Im}\,g=0.
\end{equation}
It is easy to show that this condition is always fulfilled if the problem is treated within a self-consistent Hartree-Fock approach. In contrast, the Josephson current is only conserved for symmetric gaps ($\Delta_L=\Delta_R$) if the self-energy is computed from the fRG flow equations (\ref{eq:methods.frg.flowS}) and (\ref{eq:methods.frg.flowSD}). This can be seen by using $g^\la=\Sigma_\Delta^\la/\sum_s\sum_{i\nu}a_s(i\nu)$ to describe the flow of the off-diagonal part of the self-energy. The flow equation is determined by Eq.~(\ref{eq:methods.frg.flowSD}),
\begin{equation}
\partial_\la g^\la = \frac{U^\la}{\pi}\left(\frac{g^\la}{D^\la(i\la)} - 
\frac{\sum_s\sum_{i\omega}a_s(i\omega)\delta_{\omega,\la}}{\sum_s\sum_{i\omega}a_s(i\omega)}\right).
\end{equation}
The second term on the r.h.s.~is real only for $\Delta_L=\Delta_R$. Hence, within the fRG framework the current is conserved if and only if the superconducting gaps of the left and right lead are equal.

Further insight into the structure of the self-energy can be gained by requiring Eq.~(\ref{eq:currentconservation.se}) to hold term by term. This is only possible if
\begin{equation}\label{eq:currentconservation.setermbyterm}
\Sigma_\Delta(i\omega) = f(i\omega)\tilde\Delta(i\omega), ~~~f(i\omega)=f(-i\omega)\in\mathbb{R}.
\end{equation}
Thus, it is reasonable to consider the flow of $f^\la=\Sigma_\Delta^\la/\sum_s\Gamma_se^{i\phi_s}$ instead of $\Sigma_\Delta^\la$. For symmetric gaps ($\Delta_L=\Delta_R=\Delta$), Eq.~(\ref{eq:methods.frg.flowSD}) gives
\begin{equation}
\partial_\la f^\la = \frac{U^\la}{\pi D^\la(i\la)}\left(f^\la-\frac{\Delta}{\sqrt{\la^2+\Delta^2}}\right).
\end{equation}
Since the r.h.s.~is real so is the function $f^\la$, and $\Sigma_\Delta^\la$ is of the form (\ref{eq:currentconservation.setermbyterm}). Hence, for symmetric SC gaps the identity (\ref{eq:currentconservation.se}) is fulfilled term by term within the fRG approach.



\begin{thebibliography}{*}
\bibitem{tkdelta}
One should note that the phase boundary is only roughly (and not solely) determined by the ratio $T_K/\Delta$. In particular, the phase difference $\phi$ and the impurity energy $\epsilon$ are non-trivially influencing the phase transition. This is most clearly illustrated by analytic results obtained in the limit of large $\Delta$ (see Sec.~\ref{sec:comp.atom}).
\bibitem{choiparameter}
It should be noted that the factor $A_{\Lambda}$, which appears in Eq.~(\ref{eq:A_L}), is necessary particularly when $\Lambda$ is not close to $1$. For two-channel models one has to use a rather large value for $\Lambda$ because of the numerical problems as mentioned in Sec.~\ref{sec:methods.nrg}, and the correction due to $A_{\Lambda}$ becomes sizeable. Therefore, taking this correction into account, the value $\Gamma$ used in Ref.~\onlinecite{choi} with $\Lambda=2$ can be re-interpreted, based on Eq.~(\ref{eq:A_L}), as $\Gamma \,\Rightarrow \,\Gamma A_{\Lambda=2} = 0.04D$. Here, $A_{\Lambda=2} =1.03972$, and thus $\Gamma$ that has really been examined by CLKB can be regarded as $\Gamma=0.03847D$. Hence, the ratio of $\Delta/T_K$ we used for our calculations slightly differs from the one indicated in Ref.~\onlinecite{choi}. We have chosen $U$ and $\Delta$ such that $U=0.2D$,  and $\Delta  =  0.000888D$, $0.01420D$, and $0.0888D$.
\bibitem{choierror}
In Figs.~3(a)--3(c) of Ref.~\onlinecite{choi} the Josephson current does not vanish at $\phi=\pi$ and even has a wrong sign, although this seems to be corrected partly in Ref.~\onlinecite{disputechoise2}.
\bibitem{choilambda}
One should note that the parameters used in Ref.~\onlinecite{choi} are $\la=2$ and $N_c=600$.\cite{choiNc}
\bibitem{QMC_normal}
In the case of the normal leads $\Delta=0$, the transition probability can be set up without introducing the ultraviolet cutoff for the summation over the Matsubara frequencies, as described in Ref.~\onlinecite{HirchFye}.
\end{thebibliography}
\end{document}